\documentclass[12pt]{iopart}
\usepackage{rotating}
\usepackage{graphicx}
\usepackage{latexsym}
\usepackage{amsfonts,amsthm,bm}

\newcommand{\CGf}{\ensuremath{\mathcal{C}}}
\newcommand{\PGf}{\ensuremath{\mathcal{P}}}
\newcommand{\RGf}{\ensuremath{\mathcal{R}}}
\newcommand{\ave}[1]{\langle #1 \rangle}

\begin{document}

\title{Self-avoiding walks and polygons on the triangular lattice}

\author{Iwan Jensen}
\address{
ARC Centre of Excellence for Mathematics and Statistics of Complex Systems, \\
Department of Mathematics and Statistics, \\
The University of Melbourne, Victoria 3010, Australia}

\date{\today}

\ead{I.Jensen@ms.unimelb.edu.au} 


\begin{abstract}
We use new algorithms, based on the finite lattice method of series expansion, 
to extend the enumeration of self-avoiding walks and polygons on the triangular 
lattice to length 40 and 60, respectively. For self-avoiding walks to length 
40 we also calculate series for the metric properties of mean-square end-to-end 
distance, mean-square radius of gyration and the mean-square distance of a 
monomer from the end points. For self-avoiding polygons to length 58 we calculate
series for the mean-square radius of gyration and the first 10 moments of the 
area. Analysis of the series yields accurate estimates for the connective constant of 
triangular self-avoiding walks, $\mu=4.150797226(26)$, and confirms to a high degree 
of accuracy several theoretical predictions for universal critical exponents and 
amplitude combinations.
\end{abstract}

\maketitle

\eqnobysec

\section{Introduction}

Self-avoiding walks (SAWs) and polygons (SAPs) on regular lattices are combinatorial 
problems of tremendous inherent interest as well as serving as simple models of 
polymers and vesicles \cite{MSbook,HughesV1,RensburgBook}. They are fundamental 
problems in lattice statistical mechanics. 
An {\em $n$-step self-avoiding walk} $\bm{\omega}$  is 
a sequence of {\em distinct} vertices $\omega_0, \omega_1,\ldots , \omega_n$ 
such that each vertex is a nearest neighbour of it predecessor. SAWs are
considered distinct up to translations of the starting point $\omega_0$.
We shall use the symbol $\bm{\Omega}_n$ to mean the set of all 
SAWs of length $n$. A self-avoiding polygon of length $n$ is a  $n-1$-step SAW
such that $\omega_0$ and $\omega_{n-1}$ are nearest neighbours and a closed loop
can be formed by inserting a single additional step. One is
interested in the number of SAWs and SAPs of length $n$, various metric properties
such as the radius of gyration, and for SAPs one can also ask about the 
area enclosed by the polygon. In this paper we study the following properties:

\begin{itemize}\setlength{\itemsep}{0mm}
\item[(a)] the number of $n$-step self-avoiding walks $c_n$;
\item[(b)] the number of $n$-step self-avoiding polygons  $p_n$;
\item[(c)] the mean-square end-to-end distance of $n$-step SAWs $\ave{R^2_e}_n$;
\item[(d)] the mean-square radius of gyration of $n$-step SAWs $\ave{R^2_g}_n$;
\item[(e)] the mean-square distance of a monomer from the end points of $n$-step 
SAWs $\ave{R^2_m}_n$;
\item[(f)] the mean-square radius of gyration of $n$-step SAPs $\ave{R^2}_n$; and
\item[(g)] the $k^{\rm th}$ moment of the area of  $n$-step SAPs $\ave{a^k}_n$.
\end{itemize}

The metric properties for SAWs are defined by,

\begin{eqnarray*}
\ave{R^2_e}_n =  \frac{1}{c_n} \sum_{\bm{\Omega}_n} (\omega_0 - \omega_n)^2, \\
\ave{R^2_g}_n = \frac{1}{c_n} \sum_{\bm{\Omega}_n}\left [ \frac{1}{2(n+1)^2} 
\sum_{i,j=0}^n (\omega_i - \omega_j)^2 \right ], \\
\ave{R^2_m}_n = \frac{1}{c_n} \sum_{\bm{\Omega}_n} \left [ \frac{1}{2(n+1)}
\sum_{i=0}^n \left [(\omega_0-\omega_j)^2+(\omega_n-\omega_j)^2 \right ] \right ],
\end{eqnarray*}
\noindent
with a similar definition for the radius of gyration of SAPs.

It is generally believed that the quantities listed above has the asymptotic forms
as $n \to \infty$:

\numparts
\begin{eqnarray}
c_n  =  A \mu^n n^{\gamma-1}[1 + o(1)], \label{eq:asympsaw} \\
p_n  =  B \mu^n n^{\alpha-3}[1 + o(1)], \label{eq:asympsap}  \\
\ave{R^2_e}_n  =  Cn^{2\nu}[1 + o(1)], \label{eq:asympee} \\
\ave{R^2_g}_n  =  Dn^{2\nu}[1 + o(1)], \label{eq:asymprg}\\
\ave{R^2_m}_n =  En^{2\nu}[1 + o(1)],  \label{eq:asympmd} \\
\ave{R^2}_n   =  Fn^{2\nu}[1 + o(1)],   \label{eq:asympsaprg} \\
\ave{a^k}_n   =  G^{(k)}n^{2\nu k}[1 + o(1)]. \label{eq:asympmom} 
\end{eqnarray}
\endnumparts

The critical exponents are believed to be universal in that they only depend
on the dimension of the underlying lattice.  The connective constant
$\mu$ and the critical amplitudes $A$--$G^{(k)}$  vary from lattice to lattice.
In two dimensions the critical exponents $\gamma = 43/32$, 
$\alpha =1/2$ and $\nu = 3/4$ have been predicted exactly, though 
non-rigorously, using Coulomb-gas arguments \cite{Nienhuis82a,Nienhuis84a}. 

While the amplitudes are non-universal, there are many universal amplitude
combinations. Any ratio of the metric SAW amplitudes, e.g. $D/C$ and $E/C$, 
is expected to be universal \cite{CS89}.  
Of particular interest is the linear combination
\cite{CS89,CPS90} (which we shall call the CSCPS relation)
\begin{equation} \label{eq:CSCPS}
 H \;\equiv\;
   \left( 2 +  \frac{y_t}{y_h} \right)  \frac{D}{C}
   \,-\, 2 \frac{E}{C} \,+\, \frac12,
\end{equation}
where $y_t = 1/\nu$ and $y_h = 1 + \gamma/(2\nu)$. In two dimensions  
Cardy and Saleur \cite{CS89} (as corrected by 
Caracciolo, Pelissetto and Sokal \cite{CPS90}) have predicted, using 
conformal field theory, that $H = 0$. This conclusion has been confirmed by 
previous high-precision Monte Carlo work \cite{CPS90} as well as by series 
extrapolations \cite{GY90}.

Privman and Redner \cite{PR85b} proved that the combination $BC/\sigma a_0$ 
is universal, Cardy and Guttmann \cite{CG93} proved that 
$BF=\frac{5}{32\pi^2}\sigma a_0$, and Cardy and Mussardo \cite{CM93}
proved that $C/F$ is universal and gave the first theoretical estimate
of the value $C/F\approx 13.70$. 
$\sigma$ is an integer constant such that 
$p_n$ is non-zero when $n$ is divisible by $\sigma$. So $\sigma=1$ for the 
triangular lattice and 2 for the square and honeycomb lattices. $a_0$ is the 
area per lattice site and  $a_0=1$ for the square lattice, $a_0=3\sqrt{3}/4$ 
for the honeycomb lattice, and $a_0=\sqrt{3}/2$ for the triangular lattice. 

Richard, Guttmann and Jensen \cite{RGJ01} conjectured the exact form of
the critical scaling function for self-avoiding polygons and consequently 
showed that the amplitude combinations $G^{(k)}B^{k-1}$ are universal and
predicted their exact values. The exact value for $G^{(1)}=\frac{1}{4\pi}$
had previously been predicted by Cardy \cite{JLC94a}. The validity of this 
conjecture was recently confirmed  numerically to a high degree of accuracy
using exact enumeration data for SAPs on the square, honeycomb, and triangular 
lattices \cite{RJG03}. 

The asymptotic form (\ref{eq:asympsaw}) only explicitly gives the leading contribution. 
In general one would expect corrections to scaling so, e.g, 
\begin{equation}
c_n= A\mu^n n^{\gamma-1}\left [1 + \frac{a_1}{n}+\frac{a_2}{n^2}+\ldots
+ \frac{b_0}{n^{\Delta_1}}+\frac{b_1}{n^{\Delta_1+1}}+\frac{b_2}{n^{\Delta_1+2}}+\ldots
\right]
\end{equation}
In addition to ``analytic'' corrections to scaling of the form $a_k/n^k$,
there are ``non-analytic'' corrections to scaling of the form
$b_k/n^{\Delta_1+k}$, where the correction-to-scaling exponent $\Delta_1$ 
isn't an integer. In fact one would expect a whole sequence of
correction-to-scaling exponents $\Delta_1 < \Delta_2 \ldots$, which
are both universal and also independent of the observable, that is,
the same for $c_n$, $p_n$, and so on. 
In a recent paper \cite{CGJPRS} we study the amplitudes and the
correction-to-scaling exponents for two-dimensional SAWs,
using a combination of series-extrapolation and Monte Carlo methods.
We enumerated all self-avoiding walks up to 59 steps on the square lattice,
and up to 40 steps on the triangular lattice, measuring the
metric properties mentioned above, and then carried out a detailed
and careful analysis of the data in order to accurately estimate the
amplitudes and correction-to-scaling exponent. The analysis provides
firm numerical evidence that $\Delta_1=3/2$ as predicted by Nienhuis 
\cite{Nienhuis82a,Nienhuis84a}.

In this paper we give a detailed account of the algorithm used to calculate
the triangular lattice series analysed in \cite{CGJPRS,RJG03}, perform some 
further analysis of the series and confirm to great accuracy the predicted 
exact values of the critical exponents, then we briefly summarise the results 
of the analysis from \cite{CGJPRS,RJG03} and finally study other amplitude 
combinations.

\section{Enumeration of self-avoiding walks and polygons \label{sec:flm}}

The use of transfer-matrix methods for the enumeration of lattice objects
has its origin in the pioneering work of Enting \cite{IGE80e} who enumerated square 
lattice self-avoiding polygons using the finite lattice method. 
The basic idea of the finite lattice method is to calculate partial generating 
functions for various properties of a given model on finite pieces, 
say $W \times L$ rectangles of the square lattice, and then reconstruct a 
series expansion for the infinite lattice limit by combining the results from 
the finite pieces. The generating function for any finite piece is calculated 
using transfer matrix (TM) techniques. 
The algorithm we use to enumerate SAWs and SAPs on the triangular lattice builds 
on this approach and more specifically our algorithm is based in large part on the 
one devised by  Enting and Guttmann \cite{EG92} for the enumeration of SAPs 
on the triangular lattice with the generalisation to SAWs following the work
of Conway, Enting and Guttmann \cite{CEG93} and using further recent enhancements
and parallelisation as described in \cite{IJ03a,IJ04a}. 

\subsection{Basic transfer matrix algorithm}

In this section we give a detailed description of the SAW algorithm and 
then briefly outline the changes required to enumerate SAPs.

\begin{figure}[h]
\begin{center}
\includegraphics{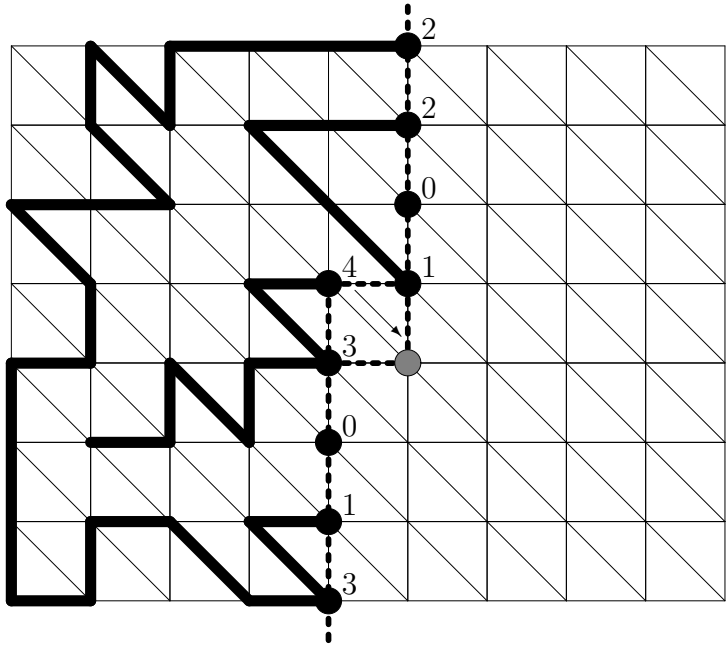}
\end{center}
\caption{\label{fig:transfer}
A snapshot of the boundary line (dashed line) during the transfer matrix 
calculation on the triangular lattice. SAWs are enumerated by successive
moves of the kink in the boundary line so that one vertex (shaded) at a time 
is added to the rectangle. To the left of the boundary line we have drawn an 
example of a partially completed SAW.}
\end{figure}

We implement the triangular lattice as a square
lattice with additional edges connecting the top-left and bottom-right
vertices of each unit cell (see fig~\ref{fig:transfer}). 
We use $W \times L$ rectangles as our finite lattices.
The most efficient implementation of the TM algorithm generally involves 
bisecting the finite lattice with a boundary (this is just a line in the 
case of rectangles) and moving the boundary in such a way as to build up 
the lattice cell by cell. The sum over all contributing graphs is 
calculated as the boundary is moved through the lattice. Due to the 
symmetry of the triangular lattice we need only consider rectangles with 
$L \geq W$. SAWs in a given rectangle are enumerated by moving the 
intersection so as to add  one vertex at a time, as shown in 
figure~\ref{fig:transfer}. In most cases it is most efficient to let the
boundary line cut through the edges of the lattice. However, on the
triangular lattice it is more efficient to let the boundary line 
cut through the vertices \cite{EG92}. Essentially this variation leads to
only half as many intersected vertices (as opposed to edges) along
the boundary line. For each configuration of occupied or empty vertices
along the intersection we maintain a generating function for partial walks 
cutting the intersection in that particular pattern. If we draw a SAW and 
then cut it by a line we observe that the partial SAW to the left of this 
line consists of a number of loops connecting two vertices (we shall refer to 
these vertices as loop-ends) in the intersection, and pieces which are connected to 
only one vertex (we call these free ends). The other end of the free piece is 
an end point of the SAW so there are at most two free ends. In addition
it is possible that the SAW touches a vertex (that is the SAW comes in along
one edge and exits along another edge both without crossing the boundary line).
All these cases are illustrated in figure~\ref{fig:transfer}.
In applying the transfer matrix technique to the enumeration of SAWs we regard 
them as sets of edges on the finite lattice with the properties:
\begin{itemize}
\item[(1)] A weight $u$ is associated with an occupied edge. 
\item[(2)] All vertices are of degree 0, 1 or 2.
\item[(3)] There are at most two vertices of degree 1 and the final graph
has exactly two vertices of degree 1 (the end points of the SAW).
\item[(4)] Apart from isolated sites, the final graph has a single connected
component.
\item[(5)] Each graph must span the rectangle from left to right and from bottom to top.
\end{itemize}

We are not allowed to form closed loops, so two loop-ends can only be joined 
if they belong to different loops. To exclude loops which close on themselves 
we need to label the occupied vertices in such a way that we can easily determine 
whether or not two loop-ends belong to the same loop. The most obvious choice 
would be to give each loop a unique label. However, on two-dimensional 
lattices there is a more compact scheme relying on the fact that two loops 
can never intertwine. Each end of a loop is assigned one of two labels 
depending on whether it is the lower end or the upper end of a loop. Each 
configuration along the boundary line can thus be represented by a set of 
states $\{\sigma_i\}$, where

\begin{equation}\label{eq:states}
\sigma_i  = \left\{ \begin{array}{rl}
0 &\;\;\; \mbox{empty vertex},  \\ 
1 &\;\;\; \mbox{vertex is a lower loop-end}, \\
2 &\;\;\; \mbox{vertex is an upper loop-end}, \\
3 &\;\;\; \mbox{touched (degree 2) vertex}, \\
4 &\;\;\; \mbox{vertex is a free end}. \\
\end{array} \right.
\end{equation}
\noindent
If we read from the bottom to the top, the configuration along the 
intersection of the partial SAW in figure~\ref{fig:transfer} is $\{310341022\}$.

\subsubsection{Updating rules}

\begin{figure}
\begin{center}
\includegraphics[scale=0.83]{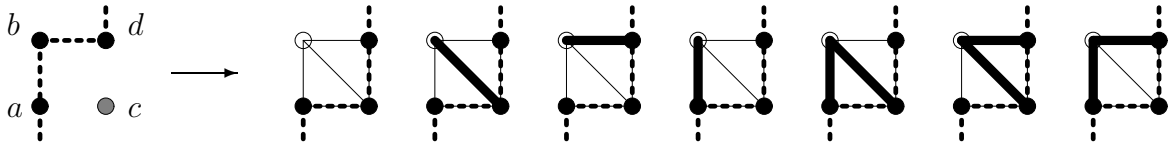}
\end{center}
\caption{\label{fig:TMit}
The seven possible outputs from a single iteration of the TM algorithm.
Depending on the states of the three vertices $a$, $b$, and $d$ in the 
input some of the outputs cannot occur. }
\end{figure}

In figure\ref{fig:TMit} we have illustrated what can happen locally as
the boundary line is moved. Before the move, the boundary line intersects
the vertices $a$, $b$ and $d$ and after the move the vertices $a$, $c$ and
$d$ are intersected by the boundary line. We shall refer to the boundary
line configuration prior to a move as the `source' and after the move 
as the `target'. In a basic iteration step we can insert bonds along the edges 
emanating from vertex $b$. Since vertex $b$ can't have degree greater than 2 we 
can at most insert two new bonds. However, depending on the states of vertices 
$a$ and $d$ in the source, some of the edge configuration in figure~\ref{fig:TMit}
may be forbidden. The updating of the partial generating function 
depends most crucially on the state of vertex $b$ and to a somewhat lesser
extent on the states of the vertices $a$ and $d$. The basic limitation
on the allowed outputs are that conditions (2)--(4) must be enforced.
In the following we shall briefly describe how the updating rules 
are derived.

\begin{description}

\item{\bf State of vertex $\bm b$ is 0.} Since vertex $b$ is empty all the outputs
in figure~\ref{fig:TMit} are possible. In the first output we insert no bonds.
This is always allowed and no changes are made to the configuration.

In the next three outputs we insert a single bond. This makes vertex $b$
of degree one and is thus only allowed if there is at most one free end in
the source. There are further restrictions on the insertion of 
a bond to vertices $a$ or $d$. Firstly if a vertex is touched (in state 3)
we cannot insert a bond since this would result in a vertex of degree 3.
Secondly if the vertex is a free end (in state 4) we join two free ends.
This leads to the formation of a completed sub-graph and
is only permitted if the resulting graph is a valid SAW. So the configuration
cannot contain other pieces of a SAW  and the only permissible states of
other vertices in the intersection are 0 and 3. If a valid SAW is created we
multiply the source generating function by $u$ (representing the new bond)
before adding it to the total for the SAW generating function. 

In the last three outputs we insert a partial loop.  Again there are 
restrictions on the insertion of bonds to vertices $a$ and $d$. 
As before we cannot insert a bond to a vertex in state 3. Otherwise
the first two outputs are always allowed. The last output is a little
more complicated. If both vertices $a$ and $d$ are in state 4  
we join two free ends and as before we check if the result is a valid
SAW and if so add this partial generating function the SAW generating function 
(this time we multiply the source generating function by $u^2$). 
If vertex $a$ is in state 1 and vertex $d$ in state 2 we cannot join the 
two vertices since this would result in a closed loop.

After the insertion of new bonds we have to assign a state to vertex $c$ and quite 
possibly change the states of vertices $a$ and $d$ (and perhaps the states of some 
other vertices in the target configuration). The state of vertex $c$ will be
0 (no bond), 1 (lower loop-end), 2 (upper loop-end) or 4 (free end). Next we
consider what happens to vertices $a$ and $d$.  When these vertices are empty in 
the source they can take the values just listed above in the target. If
they are occupied in the source they either retain their state in the target 
(no bonds inserted) or change to state 3 (a bond is inserted). In the latter
case we may have to change the state of other vertices in the target.
If we insert a free end and it joins a lower (upper) loop-end we must
change the matching upper (lower) loop-end to a free end. Otherwise
we may join two lower (upper) loop-ends and then we must change the
matching upper (lower) loop-end of the inner most loop to the
lower (upper) loop-end of the new joined loop.

\item{\bf State of vertex $\bm b$ is 1.}  A lower end of a loop enters
vertex $b$. If we insert no further bonds a new degree 1 vertex is created.
As above this is only allowed provided the source has at most one
free end. The matching upper loop-end becomes a free end. Otherwise the lower end 
has to be continued by inserting a single bond (partial loops cannot be inserted 
since this would make vertex $b$ of degree 3) either to vertex $c$ which
becomes a state 1 vertex; to vertex $a$ if not in state 3 or state 2 
(closed loop would be formed); or to vertex $d$ if not in state 3.
Again we have to change the states of vertices $a$ and $d$ when a
bond is inserted on these vertices. If the source state of the vertices was 0 the 
target state becomes 1, otherwise the target state becomes 3 and as above
we may need to change the state of other vertices as well.

\item{\bf State of vertex $\bm b$ is 2.} An upper end of a loop enters
vertex $b$. If we terminate the loop-end a new degree 1 vertex is created.
Again this is only allowed provided the source has at most one
free end. The matching lower end of the loop becomes a free end. The upper end can 
always be continued to vertex $c$; to vertex $d$ if it is not in state 3; and
to vertex $a$ provided it is not in state 3 or 1 (this would result in a
closed loop). The state of the target vertices are changed as described above.

\item{\bf State of vertex $\bm b$ is 3.} This is the simplest situation.
Vertex $b$ is of degree 2 so no bonds can be inserted and only the
output with all empty edges is allowed. The state of vertex $c$ is 0 and
the states of all other vertices are unchanged.

\item{\bf State of vertex $\bm b$ is 4.} A free end is entering vertex $b$.
If we insert no further bonds a partial walk is terminated at the vertex.
This is only allowed if the resulting graph is a valid SAW and the source 
generating function is added to the SAW generating function. The free end can 
always be continued to vertex $c$ and to vertices $a$ and $d$ if they are not 
in state 3. As before, if we join two free ends we check if it is a valid SAW 
and then add the partial generating function (multiplied by $u$) 
to the SAW generating function. Otherwise the target configuration 
is updated as described previously.

\end{description}

\paragraph{SAP updating rules.}

The updating rules used when enumerating SAPs are essentially just a subset of 
the SAW rules. Obviously there are no degree 1 vertices in the SAP case so we 
can't insert a single bond if vertex $b$ is empty. Likewise if vertex $b$ is 
occupied we must continue the loop-end. Completed SAPs are formed by closing 
a loop (if there are no other loop-ends in the source). This happens when the 
local source configuration $\{abd\}$ is $\{120\}$ and we insert a bond from 
$b$ to $a$, $\{102\}$ and we insert a partial loop from $a$ through $b$ to $d$, 
or $\{012\}$ and we insert a bond from $b$ to $d$. 

\subsubsection{Pruning}

The use of {\em pruning} to obtain more efficient TM algorithms was used 
for square lattice SAPs in \cite{JG99}. Numerically it was found that the 
computational complexity was close to $2^{n/4}$, much better than the $3^{n/4}$ 
of the original approach. We have used similar techniques for the enumerations 
carried out for this paper. Pruning allows us to discard most of the possible 
configurations for large $W$ because they only contribute at lengths greater 
than $N_{\rm max}$, where $N_{\rm max}$ is the maximal length to which we choose 
to carry out our calculations. The value of $N_{\rm max}$ is limited by the available 
computational resources, be they CPU time or physical memory. Briefly pruning 
works as follows. Firstly, for each configuration we keep track of the 
current minimum number of steps $N_{\rm cur}$ already inserted to the left
of the boundary line in order to build up that particular configuration. 
Secondly, we calculate the minimum number of additional steps $N_{\rm add}$ 
required to produce a valid SAP or SAW. There are three contributions, namely the 
number of steps required to connect the loops and free ends, the number of 
steps needed (if any) to ensure that the SAW touches both the lower and upper 
border, and finally the number of steps needed (if any) to extend at least 
$W$ edges in the length-wise direction (remember we only need rectangles
with $L \geq W$). If the sum $N_{\rm cur}+N_{\rm add} > N_{\rm max}$ we 
can discard the partial generating function for that configuration, and of 
course the configuration itself, because it won't make a contribution to the 
count up to the  lengths we are trying to obtain. 

There are no principal differences between pruning SAWs and SAPs though
the detailed implementation is more complicated for the SAW case. We found 
it necessary to explicitly write subroutines to handle the three distinct 
cases where the TM configuration contains zero, one and two free ends,
respectively. But in all cases we essentially have to go through all the 
possible ways of completing a SAW in order to find the minimum number of 
steps required. This is a fairly straightforward task though quite time 
consuming. 

\subsubsection{Computational complexity}

The time $T(n)$ required to obtain the number of polygons or walks of length $n$
grows exponentially with $n$, $T(n) \propto \lambda^n$. For the algorithm 
without pruning the complexity can be calculated exactly. Time (and memory) 
requirements are basically proportional to a polynomial (in $n$) times the 
maximal number of configurations, $N_{\rm Conf}$, generated during a calculation. 
When the boundary line is straight we can find the exact number of configurations.
First look at the situation for SAPs when there are no free ends. 
The configurations correspond to 2-coloured Motzkin paths \cite{DV84a}, since we
can map empty and touched vertices to the two colours of horizontal steps, 
vertices in the lower state to a north-east step, and vertices in the upper
state to a south-east step. The number of such paths $M_n$ with $n$ steps is 
easily derived from the generating function \cite{DV84a}
\begin{equation}\label{eq:2-Motzkin}
M(x) = \sum_{n} M_n x^n = [1-2x-(1-4x)^{1/2}]/2x^2,
\end{equation}
\noindent
which means that $M_n \sim 4^n$ as $n\to \infty$. Note that $M_n$ slightly 
over counts $N_{\rm Conf}$ since configurations without a loop aren't permitted, 
but since there are only $2^W$ of these there is no change in the asymptotic growth. 
When the boundary line has a kink (such as in figure~\ref{fig:transfer}) $N_{\rm Conf}$ 
is no longer given exactly by $M_W-2^W$. However, it is obvious that 
$N_{\rm S}(W+1) \leq N_{\rm Conf} \leq N_{\rm S}(W)$ so we see that asymptotically 
$N_{\rm Conf}$ grows like $4^W$. Since a calculation using rectangles of widths 
$\leq W$ yields the number of SAPs up to $n=2W+1$ it follows that the complexity of 
the algorithm is $T(n) \propto \lambda^n$ with $\lambda=2$.

The number of transfer matrix configurations in the unpruned SAW algorithm is 
simply obtained by inserting 0, 1 or 2 free ends into a 2-coloured Motzkin path 
and eliminating the paths corresponding to a configurations with only empty or 
touched vertices. In this case a calculation using rectangles of widths 
$\leq W$ yields the number of SAWs up to $n=W$ it follows that the complexity of 
the algorithm is $T(n) \propto \lambda^n$ with $\lambda=4$.

The pruned algorithm is much too difficult to analyse exactly. So all we
can give is a numerical estimate of the growth in the number of configurations
with $n$. That is obtained by just running the algorithm and measuring the maximal 
number of configurations generated for different values of $n$. From the actual 
numbers it appears that for the SAP case increasing $n$ by 2 increases the number 
of configurations by slightly less than a factor of 2. This would mean that for the 
pruned SAP algorithm $\lambda_p \approx \sqrt{2}$. In the SAW case it appears that 
increasing $n$ by 4 increases the number of configuration by a factor close to 5. So 
for the pruned SAW algorithm $\lambda_p \approx \sqrt[4]{5}=1.495\ldots$. So once 
again pruning results in much more efficient algorithms.

\subsubsection{Parallelisation}

The transfer-matrix algorithms used in the calculations of the
finite lattice contributions are eminently suited for parallel
computation. The bulk of the calculations for this paper
were performed on the facilities of the Australian
Partnership for Advanced Computing (APAC). The APAC facility is an HP 
Alpha server cluster with 125 ES45's each with four 1 Ghz chips for a total of 
500 processors in the compute partition. Each server node has at least 
4 Gb of memory. Nodes are interconnected by a low latency high bandwidth 
Quadrics network. 

The most basic concern in any efficient parallel algorithm is
to minimise the communication between processors and ensure that
each processor does the same amount of work and uses the same amount 
of memory. In practice one naturally has to strike some compromise
and accept a certain degree of variation across the processors.

One of the main ways of achieving a good parallel algorithm using 
data decomposition is to try to find an invariant under the
operation of the updating rules. That is we seek to find some property
of the configurations along the boundary line which
does not alter in a single iteration.
The algorithm for the enumeration of SAPs and SAWs are quite complicated 
since not all possible configurations occur due to pruning
and an update at a given set of vertices might change the state of 
a vertex far removed, e.g., when two lower loop-ends are joined
we have to relabel one of the associated upper loop-ends as
a lower loop-end in the new configuration.
However, there is still an invariant since any vertex not
directly involved in the update cannot change from being 
empty to being occupied and vice versa, likewise a touched
vertex will remain unchanged. That is only the kink vertices 
can change their occupation or touched status. This invariant
allows us to parallelise the algorithm in such a way
that we can do the calculation completely independently on each
processor with just two redistributions of the 
data set each time an extra column is added to the lattice. 
We have already used this scheme for SAPs \cite{IJ03a} and 
lattice animals \cite{IJ03b} and refer the interested
reader to these publications for further detail. Our parallelisation
scheme is also very similar to that used by Conway and Guttmann 
\cite{CG96,GC01}.

\subsection{Metric properties and area-weighted moments}
 
The generalisation of the algorithm required to calculate metric properties and 
area-weighted moments has been described in detail in \cite{IJ00a,IJ04a} in the 
square lattice case. Only some minor adjustments are needed in order to apply 
these ideas to metric properties on the triangular lattice (no changes are needed
for the area-weighted moments). We have represented the triangular lattice as
a square lattice with extra edges along one of the main diagonals in a unit cell.
A point $(s,t)$ on the square lattice is the point $(x,y)$ on the triangular lattice 
where $x=s+\frac12 t$ and $y=\frac{\sqrt{3}}{2}t$. As shown in \cite{IJ00a,IJ04a}
calculation of metric properties involves summation over products of the $x$ and
$y$ coordinates of the distance vectors. To be explicit we define the radius of 
gyration according to the {\em vertices} of the SAW. Note that the number of 
vertices is one more than the number of steps. The radius of gyration of $n+1$ 
points at positions ${\bf r}_i$ is 

\begin{equation} 
(n+1)^2 \ave{R^2_g}_n = \sum_{i>j} ({\bf r}_i-{\bf r}_j)^2 =
n\sum_i (x_i^2+y_i^2)-2\sum_{i>j}(x_ix_j+y_iy_j).
\end{equation}

From the triangular lattice coordinates we see that both $x_i x_j$ and $y_i y_j$
carry a factor $\frac14$ so in order to ensure that we get integer coefficients
we multiply by 4 and the algorithm will thus calculate the coefficients 
$4(n+1)^2 c_n\ave{R^2_g}_n$. In order to do this we maintain five partial generating 
functions for each possible boundary configuration, namely

\begin{itemize}
\item $C(u)$, the number of (partially completed) SAWs.
\item $X^2_g(u)$, the sum over SAWs of the squared components of the 
distance vectors.
\item $X_g(u)$, the sum of the $x$-component of the distance vectors.
\item $Y_g(u)$, the sum of the $y$-component of the distance vectors.
\item $XY_g(u)$, the sum of the `cross' product of the components of the  
distance vectors, that is, $\sum_{i>j}(x_ix_j+y_iy_j)$.
\end{itemize}

As the boundary line is moved to a new position each target configuration
$S$ might be generated from several sources $S'$ in the previous boundary 
position. The partial generating functions are updated as follows, 
with $(s,t)$ being the coordinates of the newly added vertex on the square lattice:

\begin{eqnarray}\label{eq:rgupdate}
C(u,S) & = & \sum_{S'} u^{n'} C(u,S'), \nonumber \\
X^2_g(u,S) & = & \sum_{S'} u^{n'}[X^2_g(u,S')+\delta_g ((2s+t)^2+3t^2)C(u,S')],\nonumber \\ 
X_g(u,S) & = & \sum_{S'} u^{n'}[X_g(u,S)+ \delta_g (2s+t)C(u,S')], \\ 
Y_g(u,S) & = & \sum_{S'} u^{n'}[Y_g(u,S)+ \delta_g tC(u,S')], \nonumber \\ 
XY_g(u,S) & = & \sum_{S'}  u^{n'} [XY_g(u,S')+\delta_g (2s+t)X_g(u,S')
                                  +\delta_g 3t Y_g(u,S')], \nonumber 
\end{eqnarray}
\noindent
where $n'$ is the number of steps added to the partial SAW. $\delta_g=0$ if 
the new vertex is empty (has degree 0) and $\delta_g=1$ if the new vertex is 
occupied (has degree $>0$).

The updating rules for the other metric properties are generalised similarly.

\subsection{Enumeration results}

We calculated the number of polygons up to perimeter 60, while the radius of 
gyration and first 10 area-weighted moments were obtained up to perimeter 58.
We calculated the number of SAWs, their mean-square radius of gyration,
mean-square end-to-end distance, and the mean-square distance of monomers from
the end points. These quantities were obtained for walks up to length 40.
The calculations required up to 35Gb of memory using up to 40 processors at
a time and in total we used about 15000 CPU hours.

In table~\ref{tab:sapser} we list the number of SAPs and their radius of gyration
while in table~\ref{tab:sawser} we list the series for the SAW problem.
These series and those for the area-weighted moments are available at 
\verb+www.ms.unimelb.edu.au/~iwan+.

\begin{sidewaystable}
\caption{\label{tab:sapser} The number, $p_n$, of embeddings of 
$n$-step polygons on the triangular lattice and their radius of gyration.}
\small
\begin{tabular}{rrrrrr} \br 
$n$ & $p_n$ & $p_nn^2\ave{R^2}_n$ & $n$ & $p_n$ & $p_nn^2\ave{R^2}_n$ \\ 
\mr
3 & 2 & 6 & 
 32 & 2692047018699717 & 25886228326621869696 \\ 
4 & 3 & 24 & 
 33 & 10352576717684506 & 110846359749047031012 \\ 
5 & 6 & 102 & 
 34 & 39902392511347329 & 474213717578995665624 \\ 
6 & 15 & 468 & 
 35 & 154126451419554156 & 2026979522666735966994 \\ 
7 & 42 & 2172 & 
 36 & 596528356905096920 & 8657009828812246231296 \\ 
8 & 123 & 9978 & 
 37 & 2313198287784319026 & 36944420238568755696168 \\ 
9 & 380 & 45816 & 
 38 & 8986249863419780682 & 157546885404468362432148 \\ 
10 & 1212 & 208686 & 
 39 & 34969337454759091232 & 671378005865890422968520 \\ 
11 & 3966 & 944766 & 
 40 & 136301962040079085257 & 2859142640844460643187642 \\ 
12 & 13265 & 4253484 & 
 41 & 532093404471021533628 & 12168301979788445465498400 \\ 
13 & 45144 & 19046580 & 
 42 & 2080235431107538787148 & 51756227545091330753357904 \\ 
14 & 155955 & 84891654 & 
 43 & 8144154378525048003270 & 220011744770726296282498056 \\ 
15 & 545690 & 376756392 & 
 44 & 31927176350778729318192 & 934740492588407244896782986 \\ 
16 & 1930635 & 1665684774 & 
 45 & 125322778845662829008494 & 3969252848247139670605665948 \\ 
17 & 6897210 & 7338822888 & 
 46 & 492527188641409773340797 & 16846468953704095289170900908 \\ 
18 & 24852576 & 32233105398 & 
 47 & 1937931188484341585677962 & 71466199766730550647612342396 \\ 
19 & 90237582 & 141171369444 & 
 48 & 7633665703654150673637363 & 303035054640652779166447899354 \\ 
20 & 329896569 & 616694403366 & 
 49 & 30101946001283232799847562 & 1284380183482800747257353493532 \\ 
21 & 1213528736 & 2687630355198 & 
 50 & 118823919397444557546535851 & 5441398704214816650431847144246 \\ 
22 & 4489041219 & 11687756315940 & 
 51 & 469508402822449711313115200 & 23043633507948438933442640818176 \\ 
23 & 16690581534 & 50726031551790 & 
 52 & 1856933773092076293566747007 & 97548735673726189271333029096494 \\ 
24 & 62346895571 & 219753786787212 & 
 53 & 7351015093472721439659392448 & 412789876403022674873495520537906 \\ 
25 & 233893503330 & 950403133411176 & 
 54 & 29126027071450640626653986531 & 1746140617537848477455116275581178 \\ 
26 & 880918093866 & 4103923685277414 & 
 55 & 115500592701344029351721102550 & 7383765950134760244068261726914950 \\ 
27 & 3329949535934 & 17695343555964594 & 
 56 & 458398255374927436357237021173 & 31212646862418768098391776139187758 \\ 
28 & 12630175810968 & 76195720234557276 & 
 57 & 1820727406941365079260306390484 & 131899272021134280524854379727885732 \\ 
29 & 48056019569718 & 327682567452126696 & 
 58 & 7237327695683743010999188700157 & 557209110506518251250962658184410206 \\ 
30 & 183383553173255 & 1407546930663067986 & 
 59 & 28789332223533619621001538109842 &  \\ 
31 & 701719913717994 & 6039368800117995984 & 
 60 & 114602547490254934327469368968190 &  \\ 
\br
\end{tabular}
\end{sidewaystable}

\begin{sidewaystable}
\centering
\caption{\label{tab:sawser} The number, $c_n$, of embeddings of $n$-step 
self-avoiding walks on the triangular lattice and their radius of gyration, 
end-to-end distance and distance of monomers from the end-points.}
\renewcommand{\arraystretch}{1.05}
\scriptsize
\begin{tabular}{rrrrr}  \br 
  $n$  &  $c_n$ &
  $\frac16 c_n \langle R^2_e \rangle _n$ &
  $\frac16 (n+1)^2 c_n \langle R^2_g \rangle _n$  &
 $\frac16 (n+1) c_n \langle R^2_m \rangle _n$ \\ 
\mr
1  &  6  &  1  &  1  &  1  \\
2  &  30  &  12  &  22  &  17  \\
3  &  138  &  97  &  282  &  178  \\
4  &  618  &  654  &  2778  &  1476  \\
5  &  2730  &  3977  &  23305  &  10667  \\
6  &  11946  &  22624  &  175194  &  70359  \\
7  &  51882  &  122821  &  1215740  &  434708  \\
8  &  224130  &  644082  &  7939156  &  2557166  \\
9  &  964134  &  3288739  &  49422491  &  14477823  \\
10  &  4133166  &  16440648  &  295993366  &  79492861  \\
11  &  17668938  &  80783857  &  1717056604  &  425633898  \\
12  &  75355206  &  391310240  &  9697408184  &  2231674940  \\
13  &  320734686  &  1872763387  &  53533130211  &  11494836257  \\
14  &  1362791250  &  8870963422  &  289769871988  &  58310378811  \\
15  &  5781765582  &  41647686501  &  1541876281342  &  291901836462  \\
16  &  24497330322  &  194014270964  &  8081886977224  &  1444405248178  \\
17  &  103673967882  &  897639074623  &  41801262603145  &  7074419785415  \\
18  &  438296739594  &  4127904278590  &  213650877117460  &  34334678700977
\\
19  &  1851231376374  &  18879838654237  &  1080407596025856  &
165283451747722  \\
20  &  7812439620678  &  85930246593928  &  5411153165106856  &
789827267540498  \\
21   &   32944292555934   &  389382874004291  &  26865804448156781  &
3749241090582031 \\
22   &   138825972053046   &  1757383045067340  & 132328831054383256   &
17689855417349797   \\
23 & 584633909268402 & 7902553525660965 &
 647064413113509344 & 83004601828121876 \\
24 & 2460608873366142 & 35417121500633314 &
 3142945284616515512 & 387503899136724032 \\
25 & 10350620543447034 & 158241760294727837 &
 15172247917136636793 & 1800616777561080887 \\
26 & 43518414461742966 & 705008848574456242 &
 72826367061554681960 & 8330920471773661365 \\
27 & 182885110185537558 & 3132749279518281223 &
 347722481262776946768 & 38390978707292879316 \\
28 & 768238944740191374 & 13886614514918779812 &
 1652126117509776447678 & 176259763248055992656 \\
29 & 3225816257263972170 & 61415827107198652263 &
 7813839241496101017943 & 806446563482615080995 \\
30 & 13540031558144097474 & 271046328280157919578 &
 36798230598686798952874 & 3677867046530479086571 \\
31 & 56812878384768195282 & 1193838903060544883615 &
 172603075240086498030932 & 16722626138383080469074 \\
32 & 238303459915216614558 & 5248569464050058190772 &
 806559315077883801952302 & 75819788411079420147060 \\
33 & 999260857527692075370 & 23034474248167644819305 &
 3755672941408238341746325 & 342850281196290726391195 \\
34 & 4188901721505679738374 & 100925879660029490332616 &
 17429779928912903943728776 & 1546457563237807336247617 \\
35 & 17555021735786491637790 & 441524252843364233569911 &
 80636231608943399450377104 & 6958970268567678359172166 \\
36 & 73551075748132902085986 & 1928731794198995523104424 &
 371943975622752362856339418 & 31245121332848941331142166 \\
37 & 308084020607224317094182 & 8413734243045682304542891 &
 1710813401690158618688146075 & 139991577634597301110308061 \\
38 & 1290171266649477440877690 & 36655327788277288494374240 &
 7848181414990001769700643892 & 625968026891459936611240307 \\
39 & 5401678666643658402327390 & 159494618902280757690831541 &
 35911648943670829119431170002 & 2793684462154188994667777314 \\
40 & 22610911672575426510653226 & 693174559672551318610401776 &
 163929038497681452701025717812 & 12445679176337664122926617782 \\
\br
\end{tabular}
\end{sidewaystable}

\section{Analysis of the series \label{sec:analysis}}

The series studied in this paper have coefficients which grow exponentially, with 
sub-dominant term given by a critical exponent. The generic  behaviour is 
$G(u) =\sum_n g_n u^n \sim (1-u/u_c)^{-\xi},$ and hence the coefficients of the 
generating function $g_n \sim \mu^n n^{\xi-1}$, where $\mu = 1/u_c$.
To obtain the singularity structure of the generating functions we used the 
numerical method of differential approximants \cite{AJG89a}. Our main objective 
is to obtain accurate estimates for the connective constant $\mu$ and confirm 
numerically the exact values for the critical exponents $\alpha$, $\gamma$ and $\nu$. 
Estimates of the critical point and critical exponents were obtained by 
averaging values obtained from second and third order inhomogeneous 
differential approximants. The error quoted for these estimates reflects the spread 
(basically one standard deviation) among the approximants. Note that these 
error bounds should {\em not} be viewed as a measure of the true error as 
they cannot include possible systematic sources of error.

Once the exact values of the exponents have been confirmed we turn our
attention to the ``fine structure'' of the asymptotic form of the
coefficients. In particular we are interested in obtaining accurate
estimates for the amplitudes. We do this by fitting 
the coefficients to the form given by (\ref{eq:asympsaw})-(\ref{eq:asympmom}).
In this case our main aim is to test the validity of the predictions
for the amplitude combinations mentioned in the Introduction.

\subsection{Self-avoiding polygons}

The expected behaviour of the mean-square radius of gyration  (\ref{eq:asympsaprg})
and moments of area (\ref{eq:asympmom}) of SAPs results in the following predictions 
for the generating functions which we study:

\begin{eqnarray}\label{eq:saprggf}
\RGf^2_g (u)& = &\sum_{n=3}^{\infty} n^2p_n\ave{R^2}_n u^n =
    \sum_{n=3}^{\infty} r_n u^n \propto (1-u\mu)^{-(\alpha+2\nu)}, \\
\label{eq:sapamgf}
\PGf^{(k)} (u)& = &\sum_{n=3}^{\infty} p_n\ave{a^k}_n u^n =
    \sum_{n=3}^{\infty} a^{(k)}_n u^n \propto (1-u\mu)^{2-(\alpha+2k\nu)},
\end{eqnarray}
\noindent
where we have taken into account that the smallest polygon has perimeter 3.
Thus we expect these series to have a critical point at $u=u_c=1/\mu$, and as 
stated previously the exponents $\alpha=1/2$ and $\nu=3/4$.

\subsubsection{The SAP generating function}

In Table~\ref{tab:anasap} we have listed the results from our series analysis
of the SAP generating function. It is evident that the estimates for the critical 
exponent is in complete agreement with the expected value $2-\alpha=3/2$. Based on 
the estimates we find that $u_c=0.24091757(1)$. We found no evidence that the SAP
generating function had any other singularities.

\begin{table}
\caption{\label{tab:anasap} Estimates for the critical point $u_c$ and exponent 
$2-\alpha$ obtained from second and third order differential approximants to the 
triangular lattice SAP generating function. $L$ is the order of the inhomogeneous
polynomial.}
\begin{indented}
\item[]\begin{tabular}{lllll} \br
 $L$   &  \multicolumn{2}{c}{Second order DA} & 
       \multicolumn{2}{c}{Third order DA} \\ \hline 
    &  \multicolumn{1}{c}{$u_c$} & \multicolumn{1}{c}{$2-\alpha$} & 
      \multicolumn{1}{c}{$u_c$} & \multicolumn{1}{c}{$2-\alpha$} \\ \mr
 0 & 0.2409175671(28) & 1.5000142(45) & 0.2409175706(62) & 1.500006(12) \\
 2 & 0.2409175709(14) & 1.5000076(29) & 0.2409175716(30) & 1.5000071(58) \\
 4 & 0.2409175714(27) & 1.5000061(56) & 0.2409175699(40) & 1.5000078(63) \\
 6 & 0.2409175707(29) & 1.5000075(58) & 0.2409175712(29) & 1.5000065(57) \\
 8 & 0.2409175724(44) & 1.500003(10)  & 0.2409175662(80) & 1.500012(14) \\
10 & 0.2409175717(39) & 1.5000051(83) & 0.2409175704(22) & 1.5000083(41) \\
\br
\end{tabular}
\end{indented}
\end{table}

\begin{figure}
\begin{center}
\includegraphics[scale=0.6]{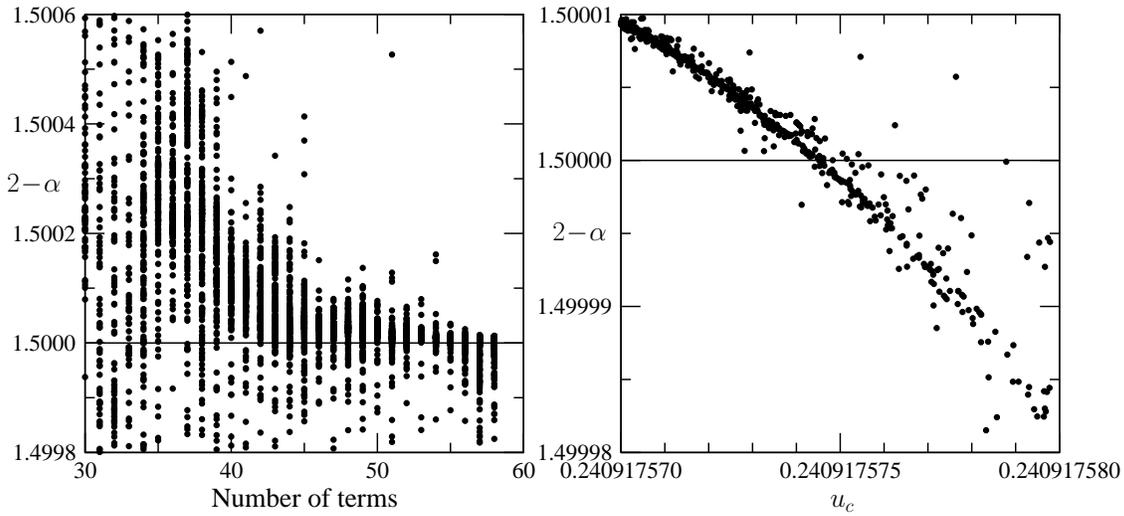}
\end{center}
\caption{\label{fig:sapexp}
Plots of estimates from third order differential approximants for 
$2-\alpha$ vs., respectively, the number of terms used by the differential 
approximants (left panel) and $u_c$  (right panel).}
\end{figure}

If we take the conjecture $\alpha =1/2$ to be true we can obtain a refined estimate 
for the critical point $u_c$. In figure~\ref{fig:sapexp} we have plotted estimates
for the critical exponent $2-\alpha$ against the number of terms used by the approximant
and against estimates for the critical point $u_c$, respectively. Each dot represents 
estimates obtained from a third order inhomogeneous differential approximant. The order 
of the inhomogeneous polynomial was varied from 0 to 10. As can be seen from the left 
panel the estimates for the critical exponent clearly include the exact value and
appear to settle down as the number of terms increases (though there is a hint of a
downwards trend). From the right panel we observe that the estimates cross the line 
$2-\alpha = 3/2$ at a value $u_c \simeq 0.2409175745$. Based on this analysis we 
adopt the value $u_c = 0.2409175745(15)$ and thus $\mu =4.150797226(26)$ as our 
final estimates.

\subsubsection{The radius of gyration}

Table~\ref{tab:anasaprg} lists the results of our series analysis for the SAP 
radius of gyration generating function. It is evident that the estimates for the 
critical exponent as obtained from third order differential approximants is in 
complete agreement with the expected behaviour. The estimates from the second order 
approximants are generally slightly lower than the expected value. One would
generally expect third order differential approximants to be superior since
they are better suited to represent complicated functional behaviour. We
take this as clear numerical confirmation  that $\nu=3/4$.

\begin{table}
\caption{\label{tab:anasaprg} Estimates for the critical point $u_c$ and exponent 
$\alpha+2\nu$ of the SAP radius of gyration generating function.}
\begin{indented}
\item[]\begin{tabular}{lllll} \br
 $L$   &  \multicolumn{2}{c}{Second order DA} & 
       \multicolumn{2}{c}{Third order DA} \\ \mr
    &  \multicolumn{1}{c}{$u_c$} & \multicolumn{1}{c}{$\alpha+2\nu$} & 
      \multicolumn{1}{c}{$u_c$} & \multicolumn{1}{c}{$\alpha+2\nu$} \\ \mr
 0 & 0.24091726(11)  & 1.99885(27) & 0.24091761(30) & 2.0001(14) \\
 2 & 0.24091727(14)  & 1.99891(26) & 0.24091728(33) & 1.99925(64) \\
 4 & 0.240917246(95) & 1.99881(14) & 0.24091713(30) & 1.99889(36) \\
 6 & 0.240917269(87) & 1.99884(14) & 0.24091741(19) & 1.99935(65) \\
 8 & 0.240917239(73) & 1.99879(11) & 0.24091743(24) & 1.99947(78) \\
10 & 0.240917281(96) & 1.99888(16) & 0.24091737(25) & 1.99932(70) \\
\br
\end{tabular}
\end{indented}
\end{table}

\subsubsection{Area-weighted moments}

In passing we only briefly mention that our analysis of the generating 
functions $\PGf^{(k)} (u)$ for area-weighted SAPs with $k\leq 10$ clearly 
confirmed the expected values, $2-(\alpha+2k\nu)$, for the critical exponents.
Suffice to say that the estimates range from $0.0005(8)$ for $k=1$ to
$-13.503(2)$ for for $k=10$.

\subsection{Self-avoiding walks}

From the expected behaviour (\ref{eq:asympsaw}) of $c_n$ and the metric properties
of SAWs (\ref{eq:asympee})-(\ref{eq:asympmd}) we get that the generating functions:

\begin{eqnarray}
\label{eq:sawgf}
\CGf (u)& = &\sum_{n=1}^{\infty} c_{n} u^n \propto (1-u\mu)^{-\gamma}, \\
\label{eq:saweegf}
\RGf^2_e (u)& = &\sum_{n=1}^{\infty}  c_{n} \ave{R^2_e}_{n}u^n 
             \propto (1-u\mu)^{-(\gamma+2\nu)}, \\
\label{eq:sawrggf}
\RGf^2_g (u)& = &\sum_{n=1}^{\infty} (n+1)^2 c_{n} \ave{R^2_g}_{n} u^n 
            \propto (1-u\mu)^{-(\gamma+2\nu+2)}, \\
\label{eq:sawmdgf}
\RGf^2_m (u)& = &\sum_{n=1}^{\infty} (n+1) c_{n} \ave{R^2_m}_{n}u^n 
              \propto (1-u\mu)^{-(\gamma+2\nu+1)}, 
\end{eqnarray}
\noindent
where the exponents $\gamma=43/32$ and $\nu=3/4$.   

\subsubsection{The SAW generating function}

In Table~\ref{tab:anasaw}  we list estimates of the critical point $u_c$  
and exponent $\gamma$ from the series for the triangular lattice SAW generating 
function. The estimates were obtained by averaging over those approximants 
which used at least the first 32 terms of the series. Based on these estimates we 
conclude that $u_c = 0.24091753(8)$ and $\gamma = 1.34368(6)$.
The estimate for $u_c$ is compatible with the more accurate
estimate $u_c = 0.2409175745(15)$ obtained above from the analysis of the SAP 
generating function. The analysis adds further support to the 
already convincing evidence that the critical exponent 
$\gamma = 43/32=1.34375$ exactly. However, we do observe that
both the central estimates for both $u_c$ and $\gamma$ are systematically
slightly lower than the expected values.

\begin{table}
\caption{\label{tab:anasaw} Estimates for the critical point $u_c$ and exponent 
$\gamma$ obtained from second and third order differential approximants to the  
square lattice SAW generating function. }
\begin{indented}
\item[]\begin{tabular}{lllll} \br
 $L$   &  \multicolumn{2}{c}{Second order DA} & 
       \multicolumn{2}{c}{Third order DA} \\ \mr
    &  \multicolumn{1}{c}{$u_c$} & \multicolumn{1}{c}{$\gamma$} & 
      \multicolumn{1}{c}{$u_c$} & \multicolumn{1}{c}{$\gamma$} \\ \mr
 0 & 0.240917491(34) & 1.343637(42) & 0.240917538(21) & 1.343687(23) \\
 2 & 0.240917529(37) & 1.343677(36) & 0.240917537(13) & 1.343686(22) \\
 4 & 0.240917529(42) & 1.343682(47) & 0.240917534(30) & 1.343682(32) \\
 6 & 0.240917524(27) & 1.343673(27) & 0.240917545(24) & 1.343693(25) \\
 8 & 0.240917523(28) & 1.343668(35) & 0.240917543(23) & 1.343692(27) \\
10 & 0.240917513(31) & 1.343662(29) & 0.240917530(22) & 1.343679(25) \\
\br
\end{tabular}
\end{indented}
\end{table}

\begin{figure}
\begin{center}
\includegraphics[scale=0.6]{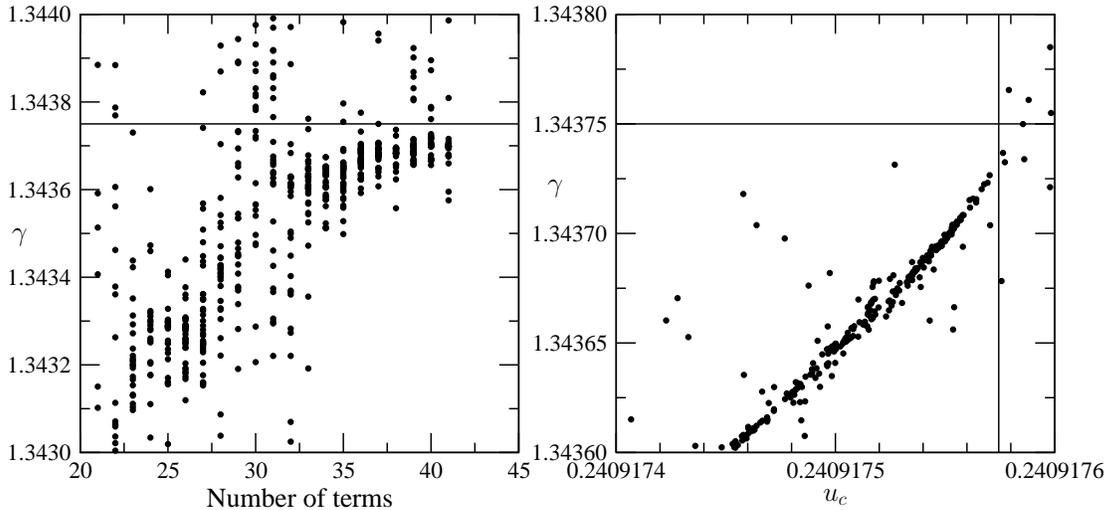}
\end{center}
\caption{\label{fig:sawexp}
Plots of estimates from second order differential approximants for 
$\gamma$ vs., respectively, the number of terms used by the differential 
approximants (left panel) and $u_c$  (right panel).}
\end{figure}

As for the SAP case we find it useful to plot the behaviour of the estimates 
for the critical exponent  $\gamma$ against both $u_c$ and the number of terms used
by the differential approximants. This is done in figure~\ref{fig:sawexp}. Each dot 
represents estimates obtained from a second order inhomogeneous differential 
approximant. From the left panel we observe that the estimates of $\gamma$ exhibits 
a certain upwards drift as the number of terms increases. So the estimates have
not yet settled at their limiting value, but there can be no doubt that the 
predicted exact value of $\gamma$ is fully consistent with the estimates. In the left 
panel we observe that the $(u_c,\gamma)$-estimates fall in a narrow range. Note that
this range does not include the intersection point between the exact $\gamma$ and 
the precise $u_c$ estimate. This is probably a further reflection of the lack of 
`convergence' to the true limiting values.

\subsubsection{The metric properties}

Finally, we turn our attention to the metric properties of SAWs. In
Table~\ref{tab:anametric} we have listed the estimates for $u_c$ and
critical exponents obtained by an analysis of the associated
generating functions (\ref{eq:saweegf})--(\ref{eq:sawmdgf}). 
The estimates from the radius of gyration $u_c=0.2409176(1)$ and
$\gamma+2\nu+2=4.84365(20)$ are in full agreement with the more
accurate SAP estimate for $u_c$ and the predicted exact exponent
value $\gamma+2\nu+2=155/32=4.84375$. The analysis of the generating functions 
for the end-to-end distance and monomer distance yield estimates of $u_c$
a little below the expected value and likewise the exponent estimates
$2.8430(5)$ and $3.8429(5)$ are a somewhat smaller that the
exact values $\gamma+2\nu+2=91/32=2.84375$ and $\gamma+2\nu+2=123/32=3.84375$,
respectively. We are fully convinced that this is because the series are
not long enough to allow the exponent estimates to settle at the true limiting 
values, as was also the case for the SAW generating function as shown in
figure~\ref{fig:sawexp}.

\begin{table}
\caption{\label{tab:anametric} Estimates for the critical point $u_c$ and critical 
exponents as obtained from third order differential approximants to the generating 
functions for the metric properties of SAWs.}
\begin{center}
\scriptsize
\begin{tabular}{lllllll} \br
 $L$   &  \multicolumn{2}{c}{$\RGf_e (u)$}  
       &  \multicolumn{2}{c}{$\RGf_g (u)$}  
       &  \multicolumn{2}{c}{$\RGf_m (u)$} \\ \hline
       &  \multicolumn{1}{c}{$u_c$} & \multicolumn{1}{c}{$\gamma+2\nu$} 
       &  \multicolumn{1}{c}{$u_c$} & \multicolumn{1}{c}{$\gamma+2\nu+2$}  
       &  \multicolumn{1}{c}{$u_c$} & \multicolumn{1}{c}{$\gamma+2\nu+1$} \\ \mr
0  & 0.240917330(86) & 2.84307(36) 
   & 0.240917594(53) & 4.843619(70)
   & 0.240917324(92) & 3.84296(17)   \\
2  & 0.240917298(62) & 2.84295(12) 
   & 0.240917600(53) & 4.843626(72)
   & 0.24091715(22)  & 3.84270(35)   \\
4  & 0.240917249(39) & 2.84272(35)
   & 0.240917605(62) & 4.843631(81) 
   & 0.24091722(23)  & 3.84281(41)   \\
6  & 0.240917311(71) & 2.84295(16)
   & 0.240917578(71) & 4.843590(99)
   & 0.24091732(17)  & 3.84299(34)   \\
8  & 0.240917328(52) & 2.842938(73)
   & 0.240917616(67) & 4.843646(89)
   & 0.240917304(43) & 3.842922(80)  \\
10 & 0.240917373(99) & 2.84303(19) 
   & 0.240917612(57) & 4.84365(10) 
   & 0.240917276(98) & 3.84285(18)   \\
\br
\end{tabular}
\end{center}
\end{table}

\subsection{Amplitudes}

\begin{figure}
\begin{center}
\includegraphics[scale=0.9]{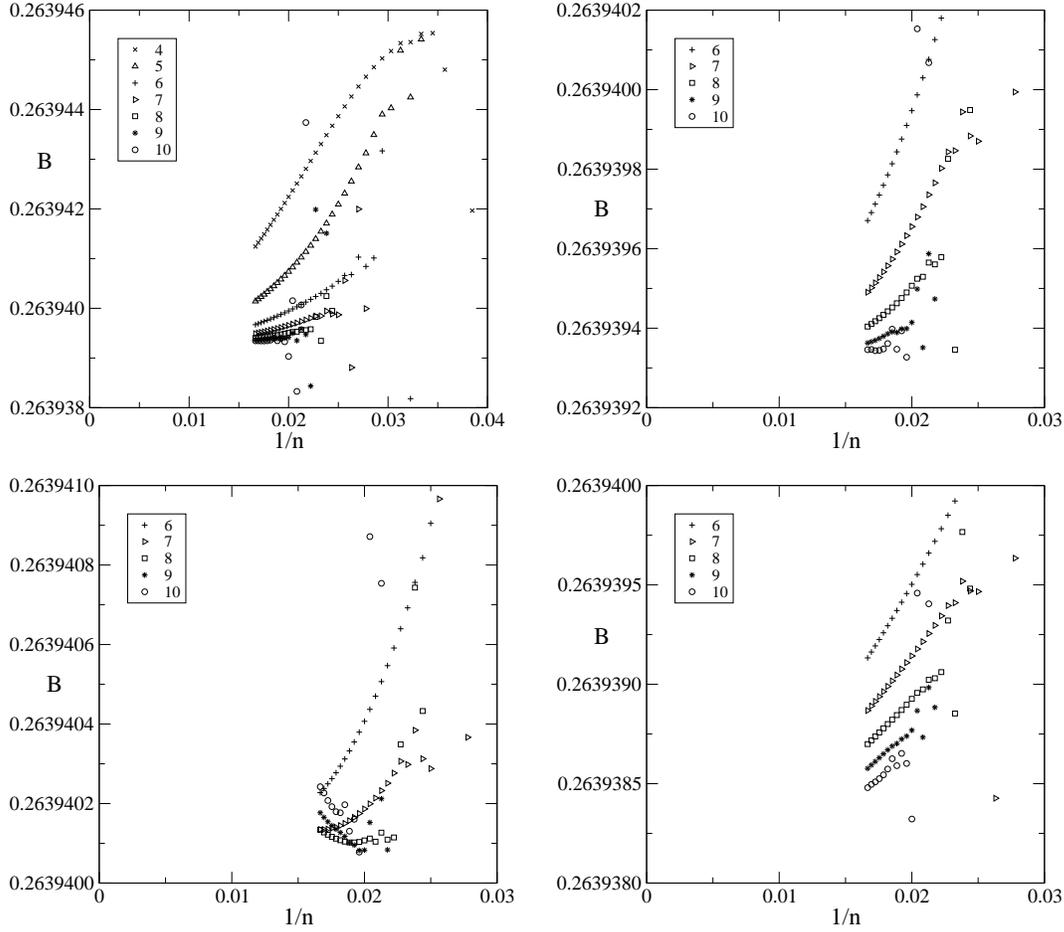}
\end{center}
\caption{\label{fig:sapgfampl}
Estimates of the leading amplitude $B$ plotted against $1/n$ using different
number of terms in the asymptotic expansion \protect{(\ref{eq:sapasymp})}. In the top 
panels we use the value $\mu=4.150797226$. The right panel just gives a more detailed 
look at the data shown in the left panel. In the bottom 
panels we use two different values $\mu=4.15079720$ (left panel) and $\mu=4.15079725$ 
(right panel) chosen to be at the extreme ends of our error-bounds for $\mu$. }
\end{figure}

The asymptotic form of the coefficients $p_n$ of the generating function of square 
lattice SAPs has been studied in detail previously \cite{CG96,JG99,IJ03a}. There is
now clear numerical evidence that the leading correction-to-scaling exponent for SAPs 
and SAWs is $\Delta_1=3/2$, as predicted by Nienhuis \cite{Nienhuis82a,Nienhuis84a}.
As argued in \cite{CG96} this leading correction term combined with the $2-\alpha=3/2$ 
term of the SAP generating function produces an {\em analytic} background term. Indeed
in the previous analysis of SAPs there was no sign of non-analytic 
corrections-to-scaling to the generating function (a strong indirect argument that 
the leading correction-to-scaling exponent must be half-integer valued). One therefore 
finds that asymptotically $p_n$ behaves as

\begin{equation}\label{eq:sapasymp}
p_n =\sim \mu^n n^{-5/2} \left [ B+ \sum_{i\geq 1} a_i/n^i \right ].
\end{equation}
\noindent
This form was confirmed with great accuracy in \cite{JG99,IJ03a}. Estimates for the 
leading amplitude $B$ can thus be obtained by fitting $p_n$ to the form 
(\ref{eq:sapasymp}) using an increasing of number of correction terms. As in 
\cite{IJ00a} we find it useful to check the behaviour of the estimates by plotting 
the results for the leading amplitude vs. $1/n$ (see figure~\ref{fig:sapgfampl}), where 
$p_n$ is the last term used in the fitting. In addition we also wish to check the 
sensitivity of the procedure to small changes in the value of $\mu$. 
Clearly the amplitude estimates in top panels are quite well converged. Notice that 
as more correction terms are added the estimates exhibit less curvature and the 
slope becomes less steep. This is very strong evidence that (\ref{eq:sapasymp}) 
indeed is the correct asymptotic form of $p_n$. The estimates shown in the bottom 
panels are not so well behaved. Those in the left panel are not monotonic and
after initially decreasing they start to increase with $n$. The estimates in 
the right panel while monotonic have much steeper slopes and the slopes do not appear
to change much as more correction terms are used. We think this is strong evidence
that  $\mu=4.150797226$ is very close to the true value. Based on the plots
in the top right panel we estimate that $B=0.2639393(1)$.

The asymptotic form of the coefficients $r_n$ in the generating function for the radius 
of gyration was studied in \cite{IJ00a}. The generating function (\ref{eq:saprggf}) has 
critical exponent $-(\alpha+2\nu)=-2$, so the leading correction-to-scaling term
no longer becomes part of the analytic background term.
We thus use the following asymptotic form: 

\begin{equation}\label{eq:rgsasymp}
r_n \sim \mu^n n \left [ BF + \sum_{i\geq 0} a_i/n^{1+i/2} \right ].
\end{equation}
\noindent
We find $BF=0.013710(1)$. This is in full agreement with the predicted exact value 
\cite{CG93} $BF=\frac{5}{32\pi^2}\sigma a_0=\frac{5\sqrt{3}}{64\pi^2}=0.013710424\ldots$,
where, for the triangular lattice, $\sigma=1$ and $a_0=\sqrt{3}/2$.
Combining the exact expression for $BF$ with the more accurate estimate for $B$
given above we estimate that $F=0.05194537(2)$.

The amplitudes of the area-weighted moments were studied in \cite{RJG03}.
We fitted the coefficients to the assumed form
\begin{equation}\label{eq:momampl}
n p_n \ave{a^k}_n \sim \mu^n n^{2k\nu+\alpha-2} k! 
\left[ G_k+\sum_{i\ge 0}a_i/n^{1+i/2} \right],
\end{equation}
where the amplitude $G_k=G^{(k)}B/k!$ is related to the amplitude
defined in equation (\ref{eq:asympmom}).
The scaling function prediction for the amplitudes $G_k$ is \cite{RGJ01}
\begin{equation}
G_{2k}B^{2k-1} = -\frac{b_{2k}}{4\pi^{3k}} \frac{(3k-2)!}{(6k-3)!}, \qquad
G_{2k+1}B^{2k} = \frac{b_{2k+1}}{(3k)!\pi^{3k+1}2^{6k+2}},
\end{equation}
where the numbers $b_k$ are given by the quadratic recursion 
\begin{equation}
b_k + (3k-4) b_{k-1} + \frac{1}{2}\sum_{r=1}^{k-1} b_{k-r}b_r=0, \qquad b_0=1.
\end{equation}
We obtained  \cite{RJG03} the results for the amplitude combinations listed 
in Table~\ref{tab:momampl}.
It is clear that the estimates for the first 10 area 
weighted moments are in excellent agreement with the predicted exact values.

\begin{table}
\caption{\label{tab:momampl}
Predicted exact values  for universal amplitude combinations and estimates
from enumeration data for square, hexagonal and triangular lattice polygons.}
\begin{center}
\scriptsize
\begin{tabular}{lllll}
\br
Amplitude  & Exact value &  Square & Hexagonal & Triangular \\
\mr
$B$  & unknown & 0.56230130(2) & 1.27192995(10) & 0.2639393(2) \\
$G_{1}$     & $0.7957747\times 10^{-1}$  & $0.795773(2)\times 10^{-1}$ 
              & $0.795779(5)\times 10^{-1}$  & $0.795765(10)\times 10^{-1}$ \\
$G_{2}B$    & $0.3359535\times 10^{-2}$  & $0.335952(2)\times 10^{-2}$ 
              & $0.335957(6)\times 10^{-2}$  & $0.335947(5)\times 10^{-2}$ \\
$G_{3}B^2$  & $0.1002537\times 10^{-3}$  & $0.100253(1)\times 10^{-3}$ 
              & $0.100255(3)\times 10^{-3}$  & $0.100251(4)\times 10^{-3}$ \\
$G_{4}B^3$  & $0.2375534\times 10^{-5}$  & $0.237552(2)\times 10^{-5}$ 
              & $0.237557(7)\times 10^{-5}$  & $0.237547(6)\times 10^{-5}$ \\
$G_{5}B^4$  & $0.4757383\times 10^{-7}$  & $0.475736(3)\times 10^{-7}$ 
              & $0.475749(10)\times 10^{-7}$  & $0.475724(15)\times 10^{-7}$ \\
$G_{6}B^5$  & $0.8366302\times 10^{-9}$   & $0.836624(5)\times 10^{-9}$ 
              & $0.836652(10)\times 10^{-9}$  & $0.83660(2)\times 10^{-9}$ \\
$G_{7}B^6$  & $0.1325148\times 10^{-10}$  & $0.132514(2)\times 10^{-10}$
              & $0.132519(5)\times 10^{-10}$  & $0.132511(5)\times 10^{-10}$ \\
$G_{8}B^7$  & $0.1924196\times 10^{-12}$  & $0.192418(2)\times 10^{-12}$
              & $0.192426(8)\times 10^{-12}$  & $0.192419(8)\times 10^{-12}$ \\
$G_{9}B^8$  & $0.2594656\times 10^{-14}$  & $0.259464(2)\times 10^{-14}$
              & $0.259472(12)\times 10^{-14}$ & $0.25948(4)\times 10^{-14}$ \\
$G_{10}B^9$ & $0.3280633\times 10^{-16}$  & $0.328062(4)\times 10^{-16}$
              & $0.328051(15)\times 10^{-16}$ & $0.32812(5)\times 10^{-16}$ \\
\br
\end{tabular}
\end{center}
\end{table}

The amplitude ratios $D/C$ and $E/C$ were estimated by direct extrapolation of the 
relevant quotient sequence, using the following method \cite{OPBG}: Given a sequence 
$\{a_n\}$ defined for $n \ge 1$, assumed to converge to a limit $a_{\infty}$
with corrections of the form $a_n \sim a_{\infty}(1 + b/n + \ldots)$,
we first construct a new sequence $\{h_n\}$ defined by $h_n = \prod_{m=1}^n a_m$.
The generating function $\sum h_n x^n \sim (1 - a_{\infty} x)^{-(1+b)}$.
Estimates for $a_{\infty}$ and the parameter $b$ can then be obtained from 
differential approximants. In this way, we obtained the estimates \cite{CGJPRS}
$D/C = 0.140296(6)$ and $E/C = 0.439649(9)$. 
These amplitude estimates leads to a high precision confirmation of the
CSCPS relation (\ref{eq:CSCPS}) $H=0.000036(34)$.

The amplitudes of the SAW generating function and the
metric properties were also studied in \cite{CGJPRS} by
fitting of the coefficients to the assumed form
\begin{eqnarray}
\label{eq:sawgfampl}
 c_n \sim \mu^n n^{\gamma-1} \left[A+\sum_{i\ge 0}a_i/n^{1+i/2} \right], \\
\label{eq:saweeampl}
 c_n\ave{R^2_e}_n/6 \sim \mu^n n^{\gamma+2\nu-1} 
\left[AC/6+\sum_{i\ge 0}a_i/n^{1+i/2} \right], \\
\label{eq:sawrgampl}
 (n+1)^2c_n\ave{R^2_g}_n/6 \sim \mu^n n^{\gamma+2\nu+1} 
\left[AD/6+\sum_{i\ge 0}a_i/n^{1+i/2} \right], \\
\label{eq:sawmdampl}
 (n+1)c_n\ave{R^2_m}_n/6 \sim \mu^n n^{\gamma+2\nu} 
\left[AE/6+\sum_{i\ge 0}a_i/n^{1+i/2} \right].
\end{eqnarray}
\noindent

\begin{figure}[h]
\begin{center}
\includegraphics[scale=0.9]{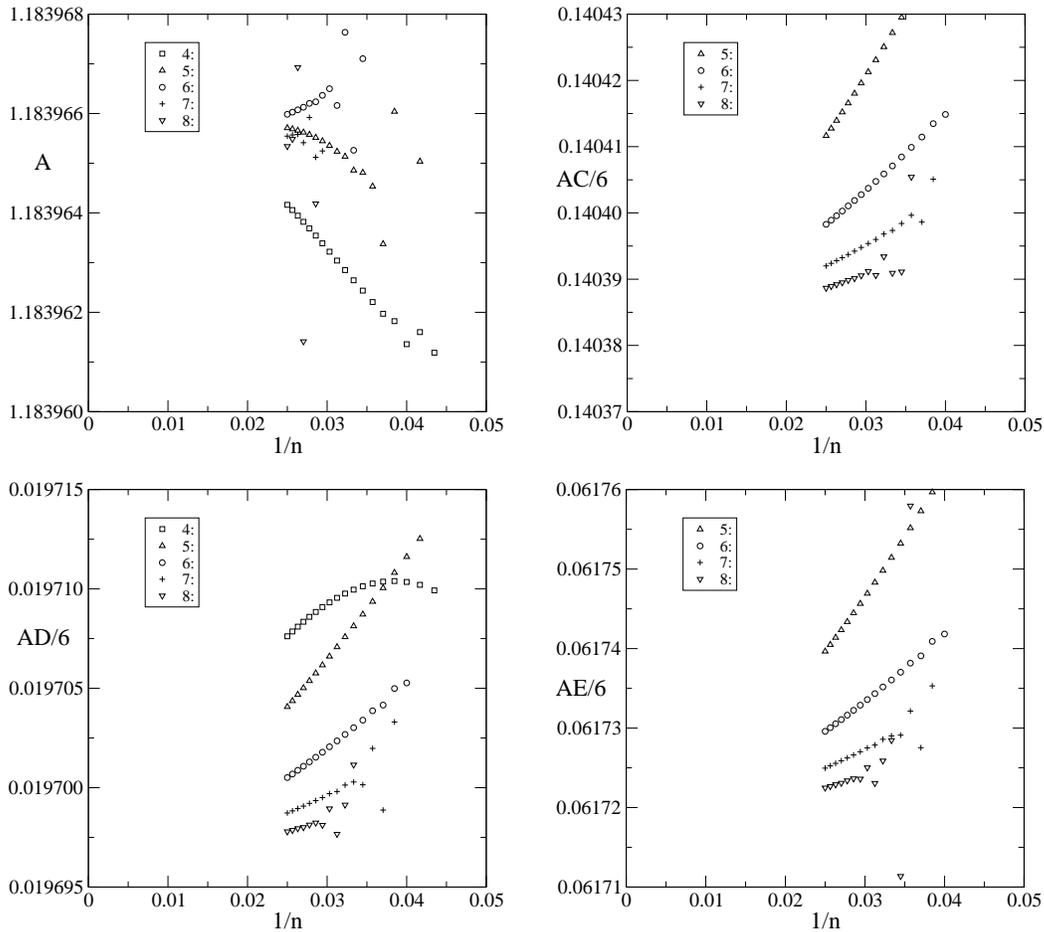}
\end{center}
\caption{\label{fig:sawampl}
Estimates of the leading amplitude $A$ for SAWs,
$AC/6$ for the end-to-end distance,
$AD/6$ for the radius of gyration,
and $AE/6$ for the monomer distance from the end-points,
plotted against $1/n$ using varying
number of terms in the asymptotic expansion.}
\end{figure}

In figure~\ref{fig:sawampl} we have plotted the estimates for the leading
amplitudes against $1/n$ while varying the number of terms used in fitting
to the expressions given above. From these we estimate that
$A=1.183966(1)$, $AC/6=0.140380(5)$, $AD/6=0.019696(3)$, and
$AE/6=0.061715(5)$. The estimate for $A$ is the same as that obtained
in  \cite{CGJPRS} while the remaining amplitude estimates are a little
lower and have smaller error-bars that those quoted in \cite{CGJPRS}.
The main reason is that here we are only interested in the
leading amplitudes and base our estimates on the fits
using 6-8 terms, while in \cite{CGJPRS} estimates for sub-leading amplitudes 
were also considered and required to be stable and consequently
only fits with up to 4 terms were considered. For the metric amplitudes
we thus obtain the estimates $C=0.71140(3)$, $D=0.099814(15)$, and
$E=0.31275(3)$. For the universal amplitude ratios we get
$D/C=0.14030(2)$ and $E/C=0.43963(5)$. We note that these estimates of 
the amplitude ratios are fully consistent with the more accurate estimates
given above. This gives us further confidence that this method for obtaining
amplitude estimates is valid. In particular, it appears, that in order to estimate
the leading amplitude, we do not have to insist that estimates for sub-leading 
amplitudes be well converged. The smaller error-bars obtained from the fits using 
6-8 terms thus appear soundly based. Naturally, some readers might wish to
take a more cautious approach.

In Table~\ref{tab:amplratio} we have summarised estimates of various universal 
amplitude combinations, obtained from the work reported in this paper and elsewhere. 
As can be seen the estimates for all lattices are in perfect agreement strongly 
confirming the universality of the various combinations.

\begin{table}
\caption{\label{tab:amplratio} Estimates of universal
amplitude combinations on some two-dimensional lattices.}
\begin{indented}
\item[]\begin{tabular}{lllll} \br
Lattice &  $D/C$ & $E/C$ & $BC/\sigma a_0$  & $H$ \\ \mr
Square \cite{CGJPRS,IJ03a} & 0.140299(6) & 0.439647(6) & 0.216835(15)&  -0.000024(28) \\
Triangular \cite{CGJPRS} & 0.140296(6) & 0.439649(9) & 0.216823(10) &  -0.000036(34)\\
Honeycomb \cite{Lin00}   & 0.1403(1) &   0.4397(2) & 0.2170(3) &  -0.00013(67) \\
Kagom\'e \cite{Lin95a,Lin99b}   & 0.140(1) & 0.440(1) & 0.2144(25) &  -0.0015(47) \\
\br
\end{tabular}
\end{indented}
\end{table}

\section{Summary and conclusion}

We have presented both improved and parallel algorithms for the enumeration 
of self-avoiding polygons and walks on the triangular lattice. These algorithms
have enabled us to obtain polygons up to perimeter length 60, their 
radius of gyration and area-weighted moments up to perimeter 58, while
for self-avoiding walks to length 40 we calculated the number of walks
as well as the metric properties of mean-square end-to-end 
distance, mean-square radius of gyration and the mean-square distance of a 
monomer from the end points. 

The analysis of the polygon series enabled us to obtain a very precise estimate for 
the connective constant $\mu=4.150797226(26)$. We confirmed to a very high degree 
of accuracy the predicted exponent values $\alpha=1/2$, $\gamma=43/32$ and $\nu=3/4$. 
We noticed that, as is the case for the square lattice problem, the SAW asymptotics is
worse behaved than the SAP asymptotics, i.e., estimates for $\mu$ and the
critical exponents are at least an order of magnitude more accurate in the SAP case.
It quite is possible that this behaviour is due to the leading correction-to-scaling
exponent $\Delta=3/2$. In the SAP case this correction simply becomes part of
the analytic background term and the SAP generating function is therefore simpler
since it only has analytic corrections to scaling.
We also obtained very accurate estimates for the leading amplitude of the
sequence $p_n$ of SAP coefficients $B=0.2639393(1)$ and using the exact expression 
for the amplitude combination $BF$ we find $F=0.05194537(2)$.
Our data for the area-weighted moments was used \cite{RJG03} to confirm
the correctness of theoretical predictions for the values of the 
amplitude combinations $G^{(k)}B^{k-1}$.
Finally we obtained accurate estimates for the critical amplitudes
$A=1.183966(1)$, $C=0.71140(3)$, $D=0.099814(15)$, and
$E=0.31275(3)$. The estimate for the ratio $C/F=13.6952(5)$ is in very 
good agreement with the theoretical estimate $C/F\approx 13.70$ \cite{CM93}.
The amplitude estimates led to a high precision confirmation of the
CSCPS relation (\ref{eq:CSCPS}) $H=0$.

\section*{E-mail or WWW retrieval of series}

The series for the problems studied in this paper 
can be obtained via e-mail by sending a request to 
I.Jensen@ms.unimelb.edu.au or via the world wide web on the URL
http://www.ms.unimelb.edu.au/\~{ }iwan/ by following the relevant links.

\section{Acknowledgments}

The calculations presented in this paper would not have been possible
without a generous grant of computer time on the server cluster of the
Australian Partnership for Advanced Computing (APAC). We also used
the computational resources of the Victorian Partnership for Advanced 
Computing (VPAC). We gratefully acknowledge financial support from 
the Australian Research Council.

\end{document}